\documentclass[10pt,letterpaper,conference]{IEEEtran}
\IEEEoverridecommandlockouts

\ifCLASSOPTIONcompsoc
  \usepackage[nocompress]{cite}
\else
  \usepackage{cite}
\fi

\ifCLASSOPTIONcompsoc
    \usepackage[caption=false, font=normalsize, labelfont=sf, textfont=sf]{subfig}
\else
\usepackage[caption=false, font=footnotesize]{subfig}
\fi

\usepackage{mathptmx}
\usepackage{amsmath,amssymb,amsfonts}
\usepackage{algorithmic}
\usepackage{mathtools}
\usepackage{xcolor}
\usepackage{multirow}
\usepackage{enumitem}
\usepackage{booktabs} %
\usepackage{array}
\usepackage{makecell}
\usepackage{xspace}
\usepackage{epsfig}
\usepackage{graphicx}
\usepackage{microtype}
\usepackage{pifont}
\usepackage[bookmarks=false]{hyperref} 
\usepackage{textcomp}
\usepackage{flushend}
\let\oldS\S
\let\orgautoref\autoref
\renewcommand{\autoref}
        {\def\figureautorefname{Fig.}%
         \def\subfigureautorefname{Fig.}%
         \def\tableautorefname{Table}%
         \def\sectionautorefname{\oldS}%
         \def\subsectionautorefname{\oldS}%
         \def\subsubsectionautorefname{\oldS}%
         \orgautoref}

\begin{document}

\title{SoK: On the Security Challenges and Risks of Multi-Tenant FPGAs in the Cloud}

\author{\IEEEauthorblockN{Shaza Zeitouni}
\IEEEauthorblockA{Technische Universit\"at Darmstadt\\
Germany\\
shaza.zeitouni@trust.tu-darmstadt.de}
\and
\IEEEauthorblockN{Ghada Dessouky}
\IEEEauthorblockA{Technische Universit\"at Darmstadt\\
Germany\\
ghada.dessouky@trust.tu-darmstadt.de}
\and
\IEEEauthorblockN{Ahmad-Reza Sadeghi}
\IEEEauthorblockA{Technische Universit\"at Darmstadt\\
Germany\\
ahmad.sadeghi@trust.tu-darmstadt.de}
\thanks{A version of this manuscript has been submitted to the IEEE for possible publication. Copyright may be transferred without notice, after which this version may no longer be accessible.}}

\maketitle
\begin{abstract}
In their continuous growth and penetration into new markets, Field Programmable Gate Arrays (FPGAs) have recently made their way into hardware acceleration of machine learning among other specialized compute-intensive services in cloud data centers, such as Amazon and Microsoft. To further maximize their utilization in the cloud, several academic works propose the spatial \emph{multi-tenant} deployment model, where the FPGA fabric is simultaneously shared among mutually mistrusting clients. This is enabled by leveraging the partial reconfiguration property of FPGAs, which allows to split the FPGA fabric into several logically isolated regions and reconfigure the functionality of each region independently at runtime.

In this paper, we survey industrial and academic deployment models of multi-tenant FPGAs in the cloud computing settings, and highlight their different adversary models and security guarantees, while shedding light on their fundamental shortcomings from a security standpoint. We further survey and classify existing academic works that demonstrate a new class of remotely-exploitable physical attacks on multi-tenant FPGA devices, where these attacks are launched remotely by malicious clients sharing physical resources with victim users. Through investigating the problem of end-to-end multi-tenant FPGA deployment more comprehensively, we reveal how these attacks actually represent only one dimension of the problem, while various open security and privacy challenges remain unaddressed. We conclude with our insights and a call for future research to tackle these challenges.
\end{abstract}


\section{Introduction}
\label{sec:intro}
Field Programmable Gate Arrays (FPGAs) are integrated circuits that can be (re)programmed after fabrication to implement custom functionalities in hardware, as opposed to application-specific integrated circuits (ASICs). FPGAs, with their more flexible computing fabric than their ASIC counterparts, higher throughput and computing power than their software counterparts, low energy consumption, and continuously increasing capacities, have been perceived to bring the best of both hardware and software worlds. 
In their steady growth and penetration of different application domains, FPGAs have also made their way into hardware acceleration of machine learning applications among other compute-intensive services. More recently, they have been increasingly adopted in data centers to accelerate cloud-based services, such as by Microsoft~\cite{catapult,brainwave}. Other enterprises, such as Amazon~\cite{amazon-f1} are even offering clients to rent FPGAs in the cloud which they can freely configure with their own logic. 

\textbf{Multi-tenant FPGAs.} To maximize FPGA utilization and the pertinent return-on-investment in cloud computing, it has been proposed to share a single FPGA fabric among the users~\cite{XMHP12,BSBG14,CSZW14,FVS15,WAHH15,KLS16,KLPW18} by leveraging the \emph{partial reconfiguration} property of FPGAs. This key distinguishing feature of FPGAs allows to reconfigure the functionality of a part, or \emph{region} of the FPGA while being deployed in-field (at runtime). 
Both reconfigurations, normal and partial, of FPGAs have led to the development of the notion of \emph{virtualized} or \emph{multi-tenant FPGAs}, which are used interchangeably to refer to FPGAs shared among several users in the cloud. While this distinction is often never clearly outlined, FPGA sharing/multiplexing/partitioning can occur either \emph{temporally} or \emph{spatially}. Temporal sharing is where the FPGA device fabric, or the accelerator configured on, is used by different users or tenants at different time slots, but never simultaneously. This is the more conventional FPGA sharing model and the one deployed now in industry solutions. More recently, however, researchers in academia and industry have been investigating spatial sharing, where different tenants' logic can be simultaneously located on different logically isolated regions of a single FPGA device. While principally possible and would further boost utilization, such a setting is not currently deployed in cloud computing architectures, and raises a plethora of new security and privacy concerns, which have not been comprehensively and systematically investigated.

\textbf{Remotely-Exploitable Physical Attacks.} However, recent academic works have investigated one dimension of this problem, where they have demonstrated how remotely-exploitable physical attacks are possible in such multi-tenant settings, with a particular focus on spatial multi-tenancy. Such attacks have been shown to compromise the availability (e.g., DoS attacks)~\cite{GOT17,MahSto19}, confidentiality (e.g., side- or covert-channel attacks)~\cite{ZhaSuh18,SGMT18,SGMT18e,RPDP18,GRE18,GRS19,TiaSze19} or integrity (e.g., fault injection attacks)~\cite{KGT18} of the victim client's logic on the FPGA. The key enabler for most of these attacks is the fact that both the victim and malicious clients have: 1)~their logic, while logically isolated, still co-located on hardware fabric that shares the supply voltage and are thus not physically isolated, 2)~and have the complete freedom to configure their allocated regions with any (malicious) hardware logic of their choice.

\textbf{End-to-End Multi-tenant FPGA Deployment.} We observe, however, that most of these works are concentrated on only one dimension of the significantly more complex and multi-dimensional problem of end-to-end deployment of multi-tenant FPGAs in the cloud. Remotely-exploitable physical attacks, while clearly a threat, are indeed not the primary security/privacy challenge stemming from these deployment settings. Other more fundamental questions remain unaddressed: How is clients' Intellectual Property (IP) protection assured? How can clients communicate their FPGA configuration files, a.k.a., bitstreams securely and privately to the device? How is secret key management on multi-tenant FPGAs possible without relying on a trusted third-party? Can current FPGA devices, which are designed to accommodate secret keys of single end-users, suffice the requirements of multi-tenant FPGAs? 

What further aggravates the problem is how these challenges and trust provisions and assumptions change for different usage and setting scenarios. Moreover, emerging FPGA-based computing trends such as data center disaggregation into resource pools~\cite{disaggreg}, disaggregation to the edge and FPGA-based confidential computing~\cite{EguVen12}, among others, further bring rise to new and different challenges. We consider that such open and fundamental challenges stand in the way of \emph{secure} deployment of multi-tenant FPGAs in the cloud.

\textbf{Contributions.} Through this work, we aim to:
\begin{itemize}[noitemsep,topsep=0pt]
\item Introduce the state of the art in current and emerging trends in multi-tenant FPGA cloud computing.
\item Survey the state-of-the-art in remotely-exploitable physical attacks in multi-tenant FPGA settings and highlight how this class of attacks represents only one dimension of the many open security and privacy challenges of end-to-end multi-tenant FPGA cloud computing.
\item Present a comprehensive list of the different security challenges and potential approaches to address them.
\item Discuss why spatial multi-tenant FPGA computing is not encouraged without certain mitigation measures in place, and conclude with a call to action for future research in addressing the challenges of end-to-end multi-tenant FPGA deployment.
\end{itemize}

\section{Background: Field Programmable Gate Arrays}
\label{sec:back}

\raggedbottom
\begin{figure*}[htbp]
\centering
\subfloat[\footnotesize Printed Circuit Board (PCB), Chips \& Dies with Reconfigurable Regions (RRs).\label{fig:board}]{
\includegraphics[width=0.55\linewidth]{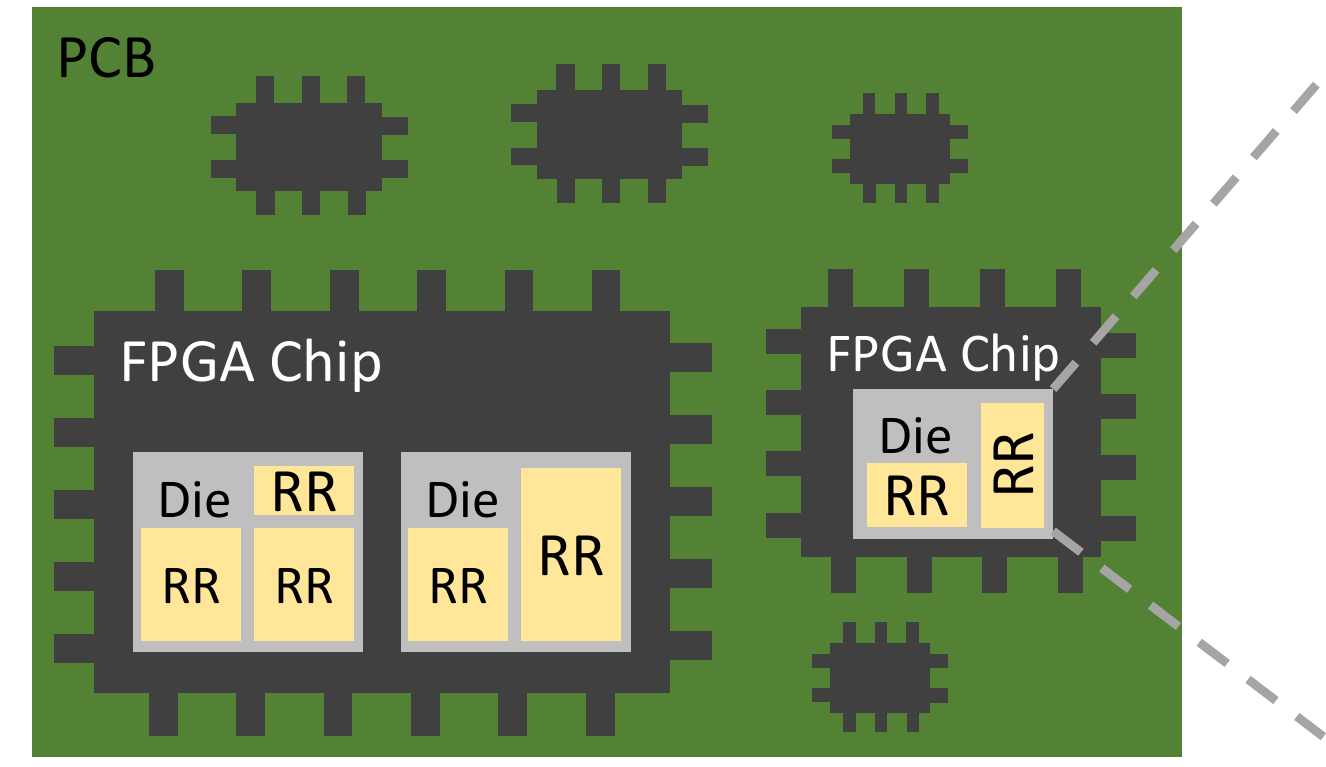}}
\subfloat[\footnotesize Basic FPGA Structure.~\cite{fpgabook}\label{fig:fpga}]{
\includegraphics[width=0.45\linewidth]{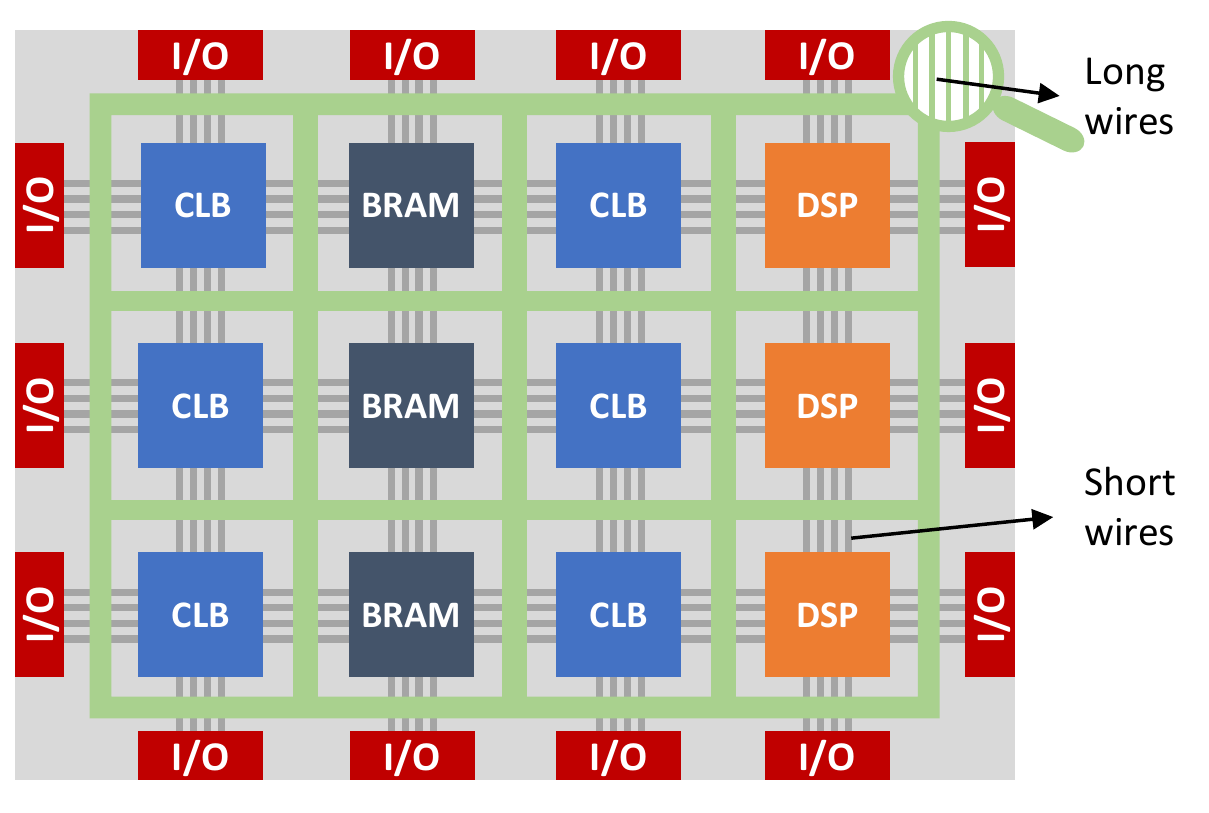}}
\caption{\small FPGAs on Board.}   
\label{fig:fpga-board}
\end{figure*}

\textbf{FPGAs} are integrated circuits that can be electrically programmed or~\emph{configured} by end users to implement different digital circuits. 
Compared to Application Specific Integrated Circuits (ASICs), FPGAs are cost-effective and have shorter time-to-market. Moreover, 
unlike ASICs, FPGAs can be reconfigured to overwrite or update an existing design.  
However, the flexible nature of FPGAs comes at additional costs in terms of area, power consumption and performance, which are mainly attributed to the programmable interconnect of FPGAs~\cite{fpgabook}.
FPGAs consist of three major components: configurable logic elements, configurable interconnects and input/output (I/O) blocks, which provide off-chip connections. 
Configurable logic blocks are connected together and to I/O blocks through the configurable routing interconnects to form the desired functionality.
Other components of FPGAs are memory blocks (BRAM) and digital signal processing blocks (DSPs) for high-performance DSP applications, as shown in~\autoref{fig:fpga}~\cite{fpgabook}. 
FPGA integration levels continue to increase with recent devices leveraging advances in 3D integration and packaging to integrate multiple dies in a single device, e.g., Intel Stratix 10 and Xilinx Virtex UltraScale+.
FPGAs are usually provided to end users on printed circuit boards (PCBs)~\autoref{fig:board}, such that a single board can have one or more FPGA chips. A single chip can pack one or multiple dies, by leveraging advances in 3D integration and packaging, as in recent FPGA chips, e.g., Intel Stratix 10 and Xilinx Virtex UltraScale+. We show this distinction, since academic proposals called for leveraging the different chips per PCB or different dies per chip in the multi-tenant FPGAs in the cloud. 
 
\textbf{Design flow} refers to the steps required to transform an abstract circuit description, written in hardware description language (HDL), such as Verilog or VHDL, into a functioning circuit on a target FPGA. 
In \emph{synthesis}, an abstract form of circuit description (functional, behavioral or structural) is translated into a functionally-equivalent gate-level representation, \emph{netlist}, using a suite of different optimizations and mapping algorithms. 
Next, in \emph{place-and-route}, the generated netlist is mapped onto the target FPGA's resources, i.e., LUTs, FFs, BRAMs, DSP cores, etc. In this step, the routing resources to connect the allocated resources are defined while preserving timing constraints, e.g., operating frequency. The developer can influence the outcome of this step by adding placement constraints as well. For example, the developer can restrict the design to be allocated on a specific region in the FPGA, or can force the routing of connections through specific channels. 
In \emph{bitstream generation}, the actual binary file that configures the target FPGA is generated.

\textbf{Partial Reconfiguration (PR)} enables dynamically reconfiguring a portion of the FPGA, while the rest of the FPGA logic continues to operate seamlessly~\cite{koch12}. An FPGA can be partitioned into a static region and one or more reconfigurable regions (RRs), such that a reconfigurable region can be configured with its own \emph{partial bitstream} without affecting other regions.
This has been one of the most significant features of FPGAs in cloud-centric applications, since it allows the cloud service provider (CSP) to partition the FPGA into one or more RRs executing different functions, thus enabling the CSP to push in new features or updates with increased flexibility and ease.
~\autoref{fig:board} shows an FPGA configuration with different RRs that can communicate with host CPU or other peripherals via pre-defined interfaces implemented in the static region. As long as the PR bitstream is accommodated by the resources of the allocated RR, the rest of the FPGA does not get affected by the dynamic reconfiguration of that region logic.

\section{FPGA Deployment in the Cloud}
\label{sec:deployment}
Early FPGAs deployments were primarily for applications such as telecommunications and ASIC prototyping. More recently, they are finding their way into large-scale data centers and cloud computing services for accelerating a wide range of compute-intensive workloads such as machine learning, genomic data processing and other scientific computations~\cite{netezza,BSBG14,CSZW14,OLQW14}. Confidential computing, another emerging trend, deploys FPGAs in the cloud to provide FPGA-based enclaves, where clients, a.k.a. \emph{tenants} can securely run (part of their) applications on sensitive data~\cite{EguVen12,EAA19}. These very different deployment settings, however, call for different requirements from the FPGA device itself as well as effective abstraction, virtualization and resource management techniques, which we briefly overview next.

\subsection{State-of-the-Art Academic Deployment Models}
\label{sec:academic}
\begin{figure}
		\centering
		\includegraphics[width=0.8\columnwidth]{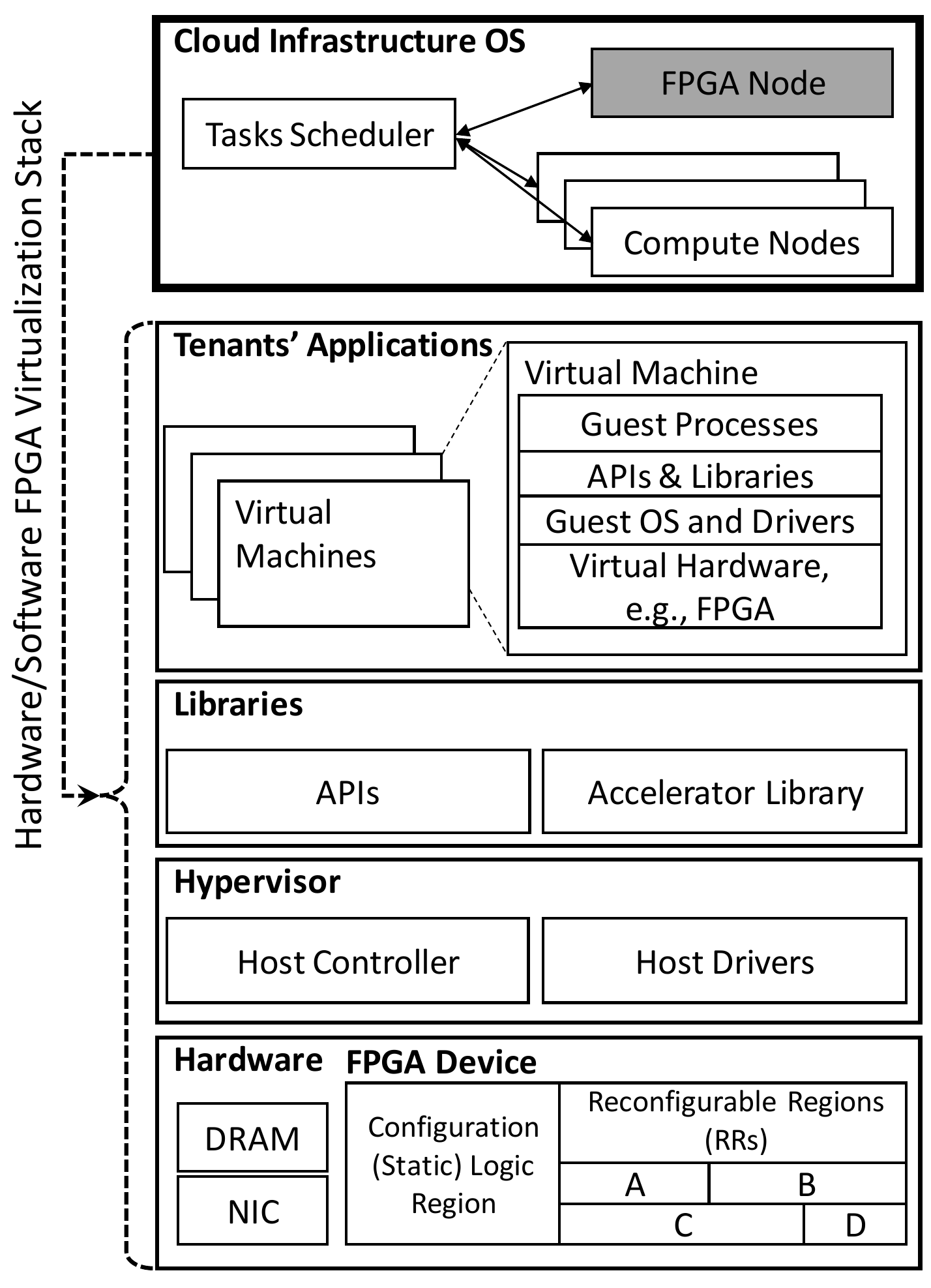}
		\caption{\small FPGA virtualization in typical cloud computing deployment.}  
		\label{fig:deploy}
	\end{figure}

In order to maximize resource utilization and return-on-investment of FPGA resources deployed in the cloud, sharing these devices among different tenants is required, which is possible due to the reconfigurable nature of FPGAs (see~\autoref{sec:back}).
To further enable this sharing, virtualization is required where the interface of interaction with the FPGAs is simplified, while abstracting away the complexities and intricacies of the underlying devices. Earlier FPGA virtualization was primarily concerned with temporal FPGA multiplexing~\cite{PlePla04} to swap in and out partitions of larger designs when device capacity was limited. Over the years, the definition of FPGA virtualization has evolved significantly (in line with evolving FPGA capacity and capability), and is now almost aligned with that of typical CPU and I/O virtualization~\cite{VPK18}, as shown in~\autoref{fig:deploy}.
Certain features about FPGAs, however, render their virtualization different from typical CPU and I/O virtualization and pose specific challenges, namely their heterogeneity, execution model and their platform/vendor-dependent configuration. 

\textbf{Virtualization Models.} Nevertheless, there has been significant progress and development in FPGA virtualization in recent years~\cite{FVS15,BSBG14,tarafdar2017}, with FPGAs now deployed in the cloud to provide either \emph{Acceleration-as-a-Service (AaaS)} or \emph{FPGA-as-a-Service (FaaS)}. The AaaS model provides tenants with FPGA-accelerated services, where the FPGA is deployed to accelerate a specific pre-defined functionality, such as a machine learning computation. The tenant cannot freely configure the FPGA hardware fabric as desired, but can only leverage its computational benefit in an acceleration service pre-defined by the cloud service provider (CSP) and the enterprise hosting the service on the cloud. On the other hand, the FaaS model enables a more flexible usage where dedicated FPGAs are allocated to tenants and freely configured by them as desired, i.e., with compatible bitstreams of their own logic. Most of the state-of-the-art work in academia is focused on boosting performance for the FaaS cloud-based model~\cite{XMHP12,BSBG14,CSZW14,FVS15,WAHH15,KLS16,KLPW18}. More details on AaaS and FaaS are in~\autoref{sec:industrial}.

Nevertheless, both cloud-based models require some mechanism of virtualization, similar to that in~\autoref{fig:deploy} to abstract hardware-specific details and represent the FPGA device and its components as yet another resource pool that can be actively managed, queried, allocated and de-allocated by tenants through the OS deployed in a cloud infrastructure. The cloud OS scheduler handles requests from tenants, creates virtual machines (VMs) for them, and schedules their tasks and resource requirements across a number of available resources and compute nodes, e.g., CPUs, memory, disks, as well as FPGAs. The tenants are thus provided access to their required resources via these VMs.

\textbf{Virtualization Granularity.} The next question of FPGA virtualization concerns the level of granularity of FPGA primitives that would get abstracted into a resource pool available for tenants to consume. Abstracting FPGAs into registers, LUTs, I/O blocks, and memory blocks for tenants to request is, however, not trivial. This is because current FPGA-based development remains largely dependent on the specific hardware fabric of the FPGA and its layout details (see~\autoref{sec:back}). These details must be visible to the FPGA toolchain in order to generate a compatible FPGA bitstream, since hardware-agnostic FPGA bitstream generation by means of overlays remains an open research problem~\cite{zuma,So2016}. Moreover, the irregular distribution of these primitives across the FPGA fabric imposes spatial constraints on how the device gets partitioned into reconfigurable regions (RRs) and how client's logic gets mapped into one of these regions. Recent FPGAs have made strides in the direction of regularity~\cite{7544760}.

Thus, FPGA virtualization schemes usually propose providing a pool of different accelerator functions from which tenants can select. 
When an accelerator function is selected, its respective pre-compiled bitstream is used to configure one of the available fixed-size reconfigurable regions (RRs) of the FPGA~\cite{CSZW14,BSBG14}, e.g. any of regions A - D in~\autoref{fig:deploy}.
These regions are each considered a ``virtual FPGA (vFPGA)'' where any accelerator can be configured on it, so long as the region provides the required resources, thus providing some degree of independence from the FPGA fabric spatial specifics~\cite{CSZW14}.
A tenant can either use an API to call an accelerator, or explicitly select a specific bitstream, or even be allowed to provide his/her own FPGA logic that gets compiled into a compatible bitstream (in case of FaaS).
Furthermore, to enable more flexibility with respect to the sizing of these RRs or vFPGAs, recent work by Khawaja et al.~\cite{KLPW18} has introduced a new layer of abstraction that encapsulates user FPGA logic and enables its mapping to dynamically sized FPGA physical zones.
Besides RRs, some logic on the FPGA device must remain statically configured (as shown in~\autoref{fig:deploy}), referred to as the FPGA shell or hypervisor logic~\cite{CSZW14}. This logic provides the physical infrastructure and logic that controls the remaining FPGA fabric (RRs A - D) and their connections to the FPGA interfaces. 

\textbf{Virtualization Stack.} To share the available FPGAs and the RRs therein across multiple tenants, the \emph{hypervisor} layer is required to provide host drivers to enable accessibility for the tenants to the underlying FPGAs. A controller module is also required to manage the resources allocation, perform address translation and provide bottom-level software interfaces to the FPGAs, thus providing multiple virtual instances of the FPGA devices~\cite{CSZW14}. \emph{Libraries} are required to wrap up the available services (different FPGA configurations) into accelerator functions and maintain their bitstreams, while also providing respective APIs for tenants to call and use the accelerators from their software applications. The FPGA device node is either configured as a stand-alone compute node and is accessed directly and independently over Ethernet~\cite{BSBG14,IAS18}, or tied to another compute node through which it can only be accessed, e.g., the host CPU through PCI Express~\cite{CSZW14}.

\textbf{FPGA (Temporal and Spatial) Multi-Tenancy.} This virtualized deployment thus enables FPGA multi-tenancy, which in its conventional definition, refers to \emph{temporal} sharing of FPGA devices. Temporal sharing is where the FPGA device itself, or the accelerator functionality configured on it, is used by different users or tenants at different time slots (in some form of time multiplexing), but never simultaneously. This FPGA sharing model is currently deployed in industry solutions. 

More recently, there has been a growing interest in both academia and industry to enable \emph{spatial} multi-tenancy on FPGA devices, where multiple tenants' logic are co-located on the same physical FPGA device simultaneously. In other words, two mistrusting tenants would have their configuration bitstreams simultaneously occupying the logically isolated RRs, e.g., A and B in~\autoref{fig:deploy}. While such a deployment model is principally possible, and would ultimately maximize resource utilization and performance gain, it raises a multitude of challenges that have not been systematically investigated before. One of the key challenges that distinguish this from CPU sharing across multiple tenants, is the security/privacy issues that arise not only at the system-level but also at the physical-level, due to the sharing of the raw hardware fabric/compute substrate itself, where the mutually mistrusting tenants have complete privilege and freedom to configure their allocated RRs as desired. Even if each tenant is restricted to configure the fabric of its allocated RR, the underlying hardware substrate and its physical implementation, such as the power supply or distribution are effectively still shared, thus isolation at this level is challenging.

\textbf{Performance vs. Security Gap.} From a performance and efficiency standpoint, there is an abundance of significant works in the literature~\cite{XMHP12,BSBG14,CSZW14,FVS15,WAHH15,KLS16,KLPW18}, among others, that have been concentrated on boosting performance and resource utilization for virtualized or multi-tenant FPGA deployment, i.e., are \emph{performance-centric}, as shown above. On the other hand, from a security and privacy standpoint, the state of multi-tenant FPGAs lags behind. Very little has been invested in investigating and enabling more \emph{security-centric} end-to-end deployment models that are suited for usage scenarios where the FPGAs are leveraged to perform security-critical tasks on sensitive data~\cite{EguVen12,HKKT17,EAA19}, or the clients are providing their proprietary designs to be accelerated on cloud FPGA devices and rightfully do not want to trust the CSP. 

Besides academia, we present next an overview of the actual deployment models adopted by the cloud computing industry to date in~\autoref{sec:industrial}. In doing so, we aim to provide readers with a comprehensive overview of the state of the art in both academia and industry with respect to FPGA virtualization, while shedding light in~\autoref{sec:security} on how multiple security and privacy challenges remain largely unaddressed.

\subsection{Commercial Deployment Solutions and Services}
\label{sec:industrial}
FPGA-accelerated computing has been investigated and recently deployed by various commercial CSPs, e.g., Microsoft, Amazon AWS, etc. In the following we briefly look into the two prominent deployment models of cloud FPGAs. 

\textbf{AaaS.} Microsoft Azure was among the first data centers to introduce FPGAs in cloud computing by augmenting CPUs with an interconnected and configurable compute layer of FPGAs to accelerate portions of large-scale software services, e.g., Bing web search~\cite{PCCC14} and host networking~\cite{FPMC18}.
Microsoft demonstrated the first proof of concept that deployed FPGAs to accelerate web search on its Bing web search engine~\cite{catapult}. 
For those original Azure deployments, Microsoft uses Intel Stratix V and Arria 10 FPGAs.
Recently, Microsoft has announced its Project Brainwave~\cite{brainwave}, a deep learning acceleration platform using Intel Stratix 10 FPGAs, aiming to deliver ``real-time AI''. The project aims to support the AI capabilities in Microsoft services like Bing web search and Skype language translations.

\textbf{FaaS.} Several CSPs offer their clients FPGA-accelerated computing instances to speed up their applications based on their needs. Prominent examples are Amazon EC2 F1~\cite{amazon-f1}, Huawei FACS FP1~\cite{huawei-facs}, 
Alibaba F1 \& F3~\cite{alibaba}, Baidu FPGA Cloud Server~\cite{baidu}, and Telekom ECS~\cite{telekom}.
The differences among them are mainly in the allocated resources (RAM, vCPUs, network, etc.) and the available FPGA devices a client can rent.
Depending on the selected instance capacity, multiple FPGA devices can be configured together to enable larger applications distributed efficiently across the FPGAs. 
Furthermore, the allotted FPGA devices are dedicated exclusively for the user during the paid duration, i.e., spatial multi-tenant FPGA usage is not yet provided, as far as we can infer from the publicly available service description.
These instances enable clients to develop FPGA-based services and accelerators by providing a VM with pre-installed and licensed FPGA design tools, i.e., Intel Quartus, Xilinx Vivado or SDAccel depending on the FPGA vendor, as well as hardware and software development kits. Ultimately, this enables both expert hardware developers and high-level language clients to benefit from these tailored instances depending on how they leverage the provided development flow.
Clients can develop their own FPGA-accelerated tasks for personal usage or for commercial purposes as in Amazon AWS or Huawei, where clients can offer their accelerators in the corresponding marketplace.  
After the hardware development process is complete and a particular FPGA image is generated, users can then proceed with the software development process, i.e., to develop, debug and run the user applications that will benefit from the FPGA-accelerated tasks.

However, one primary concern among the aforementioned solutions is IP protection for clients who run their IP, e.g., trained machine learning models on private data, on the available FPGA resources in the cloud, among many others. In light of the presented FPGA-based cloud computing solutions in both academia and industry, we outline next a more comprehensive end-to-end analysis of the different privacy and security challenges that arise.

\section{Security and Privacy Concerns}
\label{sec:security}
In such multi-layered cloud FPGA deployment settings, that are emerging in different flavors and intricacies, many fundamental security and privacy challenges arise. Particularly, for temporal and spatial FPGA multi-tenancy, new specific challenges arise that, to date, remain open and unaddressed.

\subsection{Stakeholders} 
\begin{figure}[htbp]
		\centering
		\includegraphics[width=\columnwidth]{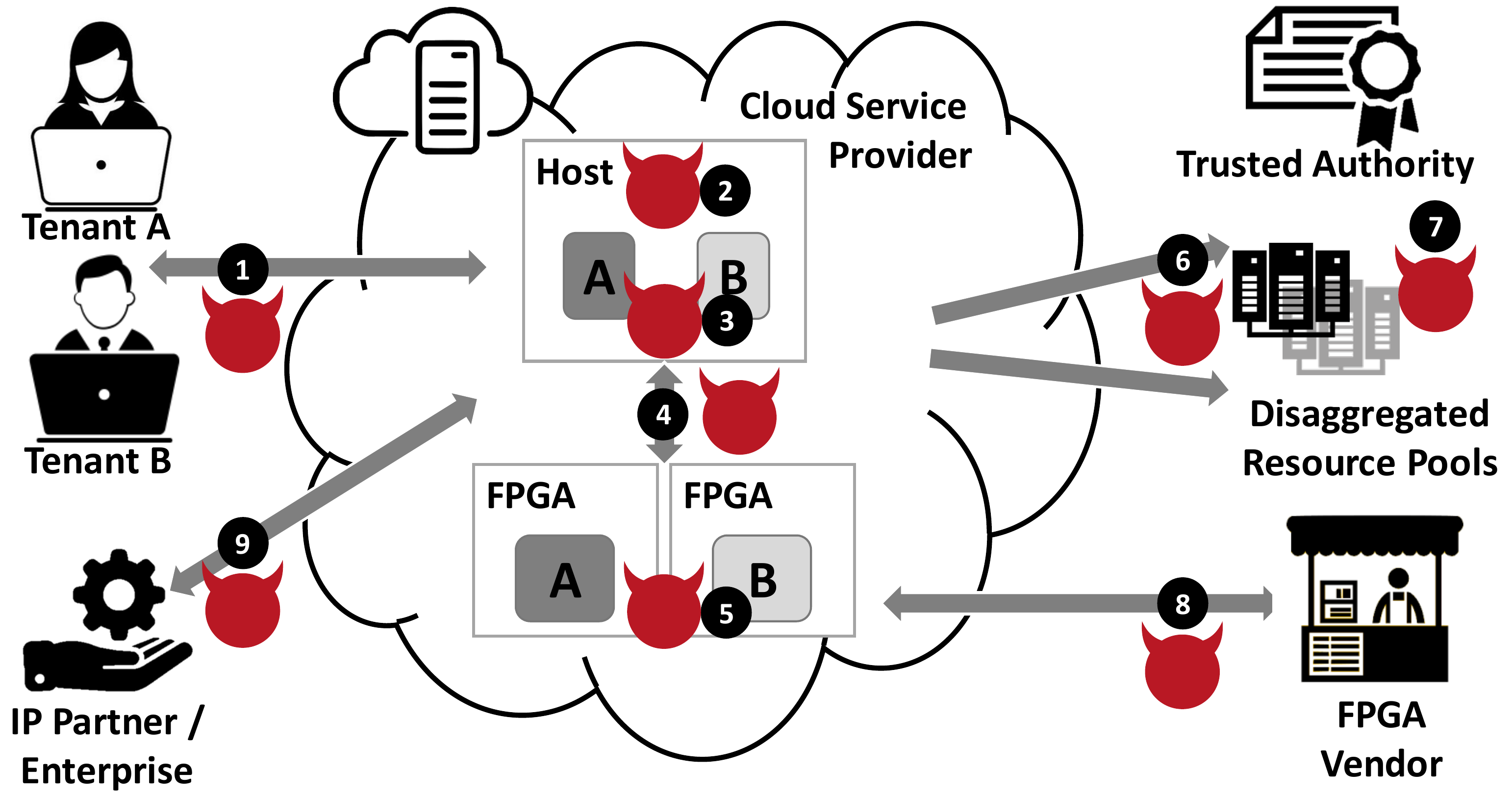}
		\caption{\small Stakeholders and their security/privacy concerns in FPGA-based cloud computing.}  
		\label{fig:security}
	\end{figure}

We enlist here the various stakeholders in FPGA cloud computing deployment in more detail. We also discuss how their different security/privacy concerns and obligations and the relationships among them both impose and influence different challenges and trust assumptions in an inevitable interplay, as abstracted in~\autoref{fig:security}.

FPGA-based cloud services are commonly deployed by a \emph{CSP}, which provides different usage models and services that involve heterogeneous architectures as discussed in \autoref{sec:deployment}. This usually involves a standard CPU-based host interfacing with co-processors and accelerator devices, such as GPUs and FPGAs, and offloading particular computation tasks to them as shown in~\autoref{fig:security}. Thus, users/clients, a.k.a. \emph{tenants} would purchase or rent the desired computation capacity and facilities and communicate their workload (which may include private/confidential data as well as intellectual property (IP) designs) to the CSP. The CSP, being the platform owner, may offload the workload from more mutually distrusting tenants to an accelerator device simultaneously, to maximize resource utilization and return on their investment, while reaping even higher performance gains. These FPGA devices and their development environments/toolchains are provided by the \emph{FPGA vendor} to the CSP, which are, in turn, made available (besides additional toolchains also provisioned by the CSP) for the tenants to use over the CSP provided service.

Besides tenants' IP, the CSP itself may be providing IP, as well as the FPGA vendor and other \emph{IP partners} or enterprises, which are developing their own third-party IP and releasing them in an IP marketplace (hosted by the CSP) for tenants to purchase and use, e.g., Amazon AWS SaaS. This could be proprietary algorithms.%
To establish secure communication channels between the users and the CSP, a \emph{trusted authority (TA)} is required for the management and provisioning of digital certificates, which could either be yet another dedicated party or the CSP itself. Moreover, if FPGA bitstream encryption (for preserving IP confidentiality) is supported, then either the FPGA vendor or the TA would have to assume the responsibilities of secure key provisioning for both the FPGA device and the tenants and digital certificates management. Further amplifying the complexity of these relationships, a very recent trend for maximizing data center efficiency and flexibility also involves the CSP disaggregating its facilities across multiple different distributed \emph{resource pools}~\cite{disaggreg}, and thus migrating the computation to these resource pools. 
 
Other stakeholders are also involved, which we purposefully keep out of scope to render this more approachable and practical. Examples include foundry facilities that hardware vendors rely on for fabricating their devices, as well as independent entities that issue certificates for CSPs' compliance with security and privacy standards. We assume, however, that trusting these entities, especially fabrication facilities (for the CPUs and FPGA devices acquired by the CSP) is a reasonable and fundamental assumption one has to make. Moreover, such threats and their proposed mitigation mechanisms occur at a very different layer of abstraction, and are an active and orthogonal research focus~\cite{TSBH20}. 

\subsection{Threat Landscape} 
With these stakeholders and multiple mutually distrusting tenants simultaneously sharing FPGA resources (either temporally or spatially) in mind, we identify a list of different (non-exhaustive) security and privacy concerns. These can be classified into either \textbf{access control}, \textbf{IP/data protection} or \textbf{physical-level threats}, and are as follows:
\begin{itemize}
	\item Tenant data must be protected against malicious network adversaries when it is sent to the CSP \ding{202}.
	\item Tenant data must be protected against potentially compromised system software \ding{203}.
	\item Tenant data must be protected against potentially compromised co-tenants' workload \ding{204}.
	\item Tenant data, which may be offloaded to the accelerator device (FPGA), must be protected when transmitted to/from the FPGA \ding{205}. This largely depends on, and may also influence, the choice of communication link between the host and the FPGA, e.g., physical link such as PCI Express, Ethernet or a wireless link. Possible attacks include man-in-the-middle attacks, payload misdirection, and device/communication spoofing or sniffing. This also includes protecting both the confidentiality and integrity of FPGA bitstreams (clients' own and enterprises' proprietary IP) before and after FPGA configuration.
	\item Tenant data must be protected from untrusted and potentially malicious or compromised tenants executing on the FPGA device. Damage, which can be caused by malicious logic executing on the FPGA, does not only affect co-tenants but also the cloud infrastructure (see~\autoref{sec:pla}) \ding{206}.
	\item In the case of disaggregation (both vertical and horizontal) to distributed resources, tenant data must be protected in transit. This again includes all network attacks \ding{207}.
	\item Given the distributed nature of such resources, tenant data must also be protected at rest and in processing. This is especially a concern with disaggregation to the edge since physical access attacks are a more likely a threat, as opposed to a single, centralized and physically secure data center \ding{208}.
	\item The CSP (and consequently the tenants) must be able to trust (or be provided with the relevant assurances/proof of trust) the FPGA devices provided by the FPGA vendor, their fabric, the security of the interfaces (particularly the debug interfaces) integrated, the cryptographic primitives baked in (which constitute part of the root of trust), and all provisioned toolchains \ding{209}.
	\item Enterprises hosting their own solutions and IP using CSP services must be protected from malicious access, counterfeiting, tampering and be able to manage, query and monitor their IP licensing rights and usage \ding{210}. 
\end{itemize}

Several of these security threats, aside from ones that arise specifically from FPGA use, have their analogues in traditional cloud-based VM computing.
Moreover, based on the cloud infrastructure, deployment settings, and assumed adversarial model for a given usage scenario, some of them may not even apply, and others may not require explicit catering and assurances, if certain reasonable trust assumptions are instead made. For example in performance-centric settings, if the tenants are willing to trust the CSP with their own data and IP, security concern~\ding{203} is no longer valid. In security-centric settings where the tenant does not trust the CSP with its security-sensitive data or its IP and demands more guarantees, all the above listed security concerns would hold and a CSP would be required to provide assurances to mitigate them. Examples include deploying a Trusted Execution Environment (TEE), e.g. Intel SGX's~\cite{sgx} or AMD's SEV~\cite{amd-sev}, to achieve isolation of tenants and OS, thus mitigating concerns~\ding{203} and~\ding{204}. Along the same line of trusting these hardware vendors for the TEE architecture, the CSP would assume trust in the FPGA vendor, the provisioned FPGA devices and toolchains, thus directly addressing concerns~\ding{209}. This would still, however, fail to provide security assurances to mitigate concerns~\ding{205} and~\ding{206}, which would require other mitigation mechanisms in place. Notably, from a CSP's standpoint, it may actual be favorable to assume a security-centric model and the tenant's mistrust, and thus, be required to provide the relevant security guarantees for the tenants. Ultimately, this would limit their liability with respect to the clients' data and IP, and would attract more customers, generating more business and profitability. With these concerns in mind, we examine next the security mechanisms currently provided by FPGA vendors in their devices.

While the security challenges of end-to-end multi-tenant FPGA deployment raised above have received very little dedicated attention in academia, a number of works have addressed some dimensions of the problem. However, none focus exclusively on all the specific challenges and needs that arise due to cloud-based FPGA deployment and FPGA sharing among multiple tenants in that setting. In what follows next, we present an overview of the state of the art in academia with respect to system-level attacks and defenses that are concentrated on \textbf{IP protection} and \textbf{access control} issues in~\autoref{sec:sla}, followed by a more detailed survey of recently reported remotely-exploitable \textbf{physical-level attacks} in~\autoref{sec:pla} and their defenses in~\autoref{sec:def}.

\section{System-Level Attacks and Defenses}
\label{sec:sla}
We present next an overview of the state-of-the-art IP protection capabilities of recent FPGA devices for cloud deployment, followed by academic solutions proposed specifically for the cloud settings. 

\subsection{IP Protection}
\label{sec:ipencrypt}
\textbf{FPGA Vendors Support.}
We briefly outline next the protection mechanisms provided by prominent FPGA vendors for IP protection, in terms of confidentiality and integrity.

\textbf{Intel} Arria 10 and Stratix 10 families incorporate built-in AES-GCM-256 and elliptic curve cryptography ECC-256/384 for bitstream decryption and authentication, respectively. The secret key can be provisioned either in a volatile memory powered with a coin-cell battery or in a one-time programmable non-volatile memory~\cite{an556}, where both can be programmed through JTAG.
Moreover, Intel Stratix 10 FPGAs and beyond provide an irreversible tampering-protection mode to enforce bitstream encryption, where JTAG, test modes as well as FPGA configuration readback~\footnote{This enables reading configuration memory content (LUTs and programmable connections) to support the implementation of error correction for reliability issues.} are disabled, thus preventing readback of unencrypted configuration data~\cite{an556,intel_sdm}.
These services are all provided by the secure device manager (SDM), a TPM-like micro-processor, which also enables secure in-field encryption key upgrades (for single sessions) for Intel Stratix 10 FPGAs, besides permanently updating the SDM locally with a new key~\cite{intel_sdm,intel_device_security}. 

\textbf{Xilinx} provides in UltraScale and UltraScale+ families built-in AES-GCM-256 and RSA-2048 for bitstream decryption and authentication, respectively. Encryption key is stored internally in either dedicated RAM, backed up by a small externally connected battery, or in a non-volatile (eFUSE) register, both of which can be locked to prevent read/write accesses. 
The encryption key can be programmed into the device through an external JTAG interface or an internal JTAG primitive, which can be further used to perform remote updates of the BBRAM key in-field~\cite{xapp1283}.
When using RSA-2048 authentication, the hash of the public key must be programmed into an eFUSE register.
When bitstream encryption is enabled, external readback of configuration memory is automatically disabled~\cite{xapp1267}. 
To protect the encryption key, Xilinx obfuscates the secret key on UltraScale and UltraScale+ families using a metalized family key stored in the silicon, since the same key is used for all devices of the same family.
Xilinx does not provide the family key as part of its toolchain, however, only distributes it to `qualified' customers~\cite{xapp1267}.

Both Intel and Xilinx FPGAs use volatile SRAM for configuration memory. Since this loses its content at power-off, an off-chip non-volatile memory is used to store the bitstream until it is loaded to the FPGA again at the next power-up. Therefore, it is necessary to provision the bitstream securely between power cycles. 

\textbf{Microsemi} provides Intrinsic-ID SRAM-PUFs in SmartFusion 2 and IGLOO 2 FPGAs, in addition to several built-in cryptographic cores (SHA-256, AES-256, ECC, etc.). 
These built-in crypto cores are not exclusive for bitstream protection; they can be used by the users in their designs as well.
Moreover, SRAM-PUFs can be used to derive FPGA-specific public and private keys for ECC, where the public key is provided to the user and can be used for ECC encryption. 
Microsemi FPGAs have two operating modes: user mode, in which the configuration, i.e., the user's design, cannot be read out, and factory test mode, used by Microsemi during the manufacturing process and disabled when shipped to the user. 
In this mode, the configuration memory, and thus the user's design, can be read out using the proper test programs. 
However, factory test mode can be re-enabled by the user by matching factory passcodes to perform failure analysis~\cite{ug0443}. 

\textbf{IP Protection in Cloud FPGAs.} In~\cite{EguVen12}, the authors propose to leverage IP protection to create a TEE on the FPGA and protect client data and computations in the cloud from an untrusted CSP. A trusted authority (TA) is introduced for provisioning of a secret key to each FPGA securely locally before deployment in the cloud. Clients send their IP to the TA (targeting a specific FPGA instance) for encryption. The client then sends the encrypted bitstream to the FPGA instance for configuration. Alternatively, public key encryption generated by the TA is proposed, such that the client can encrypt the bitstream using the public key of a given FPGA instance and configure it directly without involving the TA. While this work indirectly addresses the problem of key provisioning on FPGAs in the cloud by introducing a trusted third party, it merely transfers the trust problem with respect to key provisioning from one party (the CSP) to a third party (the TA). 

In summary, IP protection in cloud FPGAs requires vendor \emph{post-deployment} support to provision and manage FPGA-specific secret keys, thus assuring clients that they are actually communicating with the intended FPGAs. Nevertheless, the major concern with proposed solutions to date is that the CSP has no guarantees that the encrypted bitstream is free of malicious circuits (see~\autoref{sec:sensors}). 

\subsection{Access Control} 
Yazdanshenas et al.~\cite{Betz18,Betz19} attempt to address data confidentiality for communication among different mutually mistrusting regions on one FPGA and for off-chip communication. They claim that floor-planning regions on the FPGA as physically far as possible would improve physical isolation and prevent crosstalk attacks (see~\autoref{sec:pla}). While this would be true for crosstalk, it does not mitigate power leakage attacks, and further requires knowledge of the FPGA layout and adjacent wiring. To mitigate communication and data-sniffing attacks they propose to use encryption in the FPGA static configuration logic for all incoming/outgoing FPGA traffic through its interfaces. They also propose to add encryption core wrappers around regions to further encrypt inter-region communication to mitigate more sophisticated attacks where routing multiplexers can be configured by the tenant of one region to tap into the wires of another for example. However, extensive support by the FPGA vendor is required to integrate these architectural modifications into their FPGAs. The authors also propose to leverage the hardwired Network-on-Chip (NoC) that is incorporated in the recently introduced Xilinx Versal FPGAs~\cite{versal} where the NoC can be used to connect these regions via encryption-enhanced routers.

The feasibility of other inter-region attacks has been highlighted in~\cite{EKC19}, where one tenant can gain unauthorized access to another tenant's IP by launching an address-redirection attack to load the victim's bitstream from otherwise unauthorized memory locations. Alternatively, the malicious tenant can also steal IP without directly accessing the bitstream by redirecting messages between the application and the hardware accelerator task to an unauthorized malicious hardware task to disclose sensitive data. Another threat is that of task hiding where the attacker can bypass the reconfiguration manager and reconfigure the FPGA with a malicious unregistered bitstream that e.g., damages the FPGA device (DoS).

To mitigate such attacks, the authors propose to add a secure authentication module (SAM), that is external to the FPGA device (off-chip), to manage the communication between FPGA and the external world, as well as authenticate the hardware tasks and software applications, and enforce the required access control (assuming that the reconfiguration manager is compromised). 
Upon client registration in the cloud, this SAM generates a client secret key and shares it securely with the client. The client integrates the secret key in the software application and the FPGA bitstream and both are uploaded to the cloud. When a client SW-application requests an FPGA-accelerated hardware task, the requested bitstream is fetched from its storage memory and used to configure a region on the FPGA. The SAM verifies that both the SW-application and the FPGA-accelerated task belong to the corresponding legitimate client using the implanted secret key at the registration step. If authenticated, the SAM sends both components another secret key to establish encrypted data exchange between them. Thus, the established chain of trust relies on the security of assigning the secret key to the client at registration step. However, if the bitstream (with the implanted key) is not stored in protected memory or encrypted at storage, it remains vulnerable to unauthorized access by other tenants, thus compromising the security of the entire scheme.

\section{Remotely-Exploitable Physical Attacks}
\label{sec:pla}
In this section, we present state-of-the-art remotely-exploitable physical attacks (referred to as remote physical attacks for brevity) in multi-tenant FPGAs. Some of these attacks assume spatial multi-tenancy, where the clients are either spatially co-located on a single FPGA, or even on two different FPGAs sharing the same power delivery system, while others work in temporal multi-tenancy model, where clients share the same FPGA after each other.
These attacks can either target \emph{availability} (e.g., DoS attacks), \emph{confidentiality} (e.g., side- or covert-channel attacks) or \emph{integrity} (e.g., fault injection attacks) of the client's logic on the cloud FPGA.
The key enabler for most of these attacks is that victim and malicious clients can freely configure their logic on physical fabric that share the power supply. 
For a better understanding of the attacks, we provide a brief description of the power supply system of FPGAs next, as well as the primitives deployed in such attacks in~\autoref{sec:sensors} and~\autoref{sec:pwr_virus}, before classifying the attacks in~\autoref{sec:fa},~\autoref{sec:sca} and~\autoref{sec:cca}.

\textbf{Power Distribution Network (PDN)} for an FPGA consists of a voltage regulator, decoupling capacitors and a grid of interconnects, modeled as an RLC-network, to deliver the power and ground voltages to FPGA components~\cite{pdn04,ug483}.
The voltage drop $V_{drop}$ is modeled as the sum of two components a \emph{steady-state} voltage drop $IR$ and a \emph{transient} voltage drop $L di/dt$ caused by the interconnect resistance and inductance of the RLC-network, respectively.
The power consumption is a product of voltage drop and drawn current. Since voltage supplied to the FPGA device must remain fixed at a pre-defined level, changes in power demands, which vary based on the running functionality, are manifested as changes in current demands. The voltage regulator and the decoupling capacitors operate in parallel to accommodate the changes in current demand and to compensate steady-state and transient voltage drops, respectively, to try and maintain a fixed voltage level.
However, in case of sharp transient changes (in picoseconds) in current demand of the FPGA, neither the voltage regulator nor the decoupling capacitors can respond fast enough. This leads to spatial and temporal fluctuations of the voltage level in the PDN, which may result in functional failures. 

\subsection{Reconfigurable Sensors}
\label{sec:sensors}

Newer generations of reconfigurable platforms include analog built-in sensors to monitor the FPGA's supply voltage and temperature, e.g., Xilinx System Monitor (SYSMON). However, such built-in sensors cannot capture the internal changes of operating conditions in the FPGA fabric and have a low sampling frequency, e.g., SYSMON4 has a sampling rate of 0.2 MHz for the FPGA die on the Virtex UltraScale+~\cite{ug580}. 
Therefore, fine-grained on-chip sensors, which can measure the impact of physical properties (e.g., voltage variations in power distribution network, aging, etc.) of different regions on the reconfigurable logic, have been proposed~\cite{WSC09,ZicHay12}.
Two prominent examples of such on-chip sensors, which are implemented in the FPGA fabric and can perform self-measurement of their path delays, are the ring oscillator and delay-line based sensors.   
Since propagation delay in combinatorial circuits~\footnote{Combinatorial circuits consist only of wires and logic gates.} is highly affected by process, voltage and temperature (PVT) variations, a self-measurement of combinatorial propagation delays provides a means of measuring fine-grained changes in operating conditions and process variations. Nevertheless, calibration or post-processing steps might be required to suppress the undesired effects~\cite{ZicHay12,GOKT16}.

\textbf{Ring Oscillator (RO) Based Sensor.}
A classical ring oscillator (RO) consists of an AND gate (enabler) and an odd number of serially-chained inverters, where the output of the last inverter is fed back to the input of the AND gate (along with an enable signal), thus forming a \emph{combinatorial loop} circuit. 
Such combinatorial loop circuits can be detected by running design rule check (DRC), embedded in the FPGA toolchain, on the design netlist~\footnote{DRC is deployed by Amazon AWS to prevent clients' designs containing combinatorial ROs~\url{https://github.com/aws/aws-fpga/blob/master/ERRATA.md}.}. However, \emph{sequential} latch-based RO and flip-flop based RO designs can bypass DRC tests~\cite{SSNS19}. The different RO structures are demonstrated in~\autoref{app:ro}.

When a RO circuit is enabled, the RO output signal alternates between `1' and `0' continuously. The switching frequency at the RO output, i.e., the \emph{RO frequency}, is inversely proportional to the delay of a signal traversing the feedback loop through the AND gate and chained inverters/buffers. Therefore, the oscillation frequency is mainly determined by the number of inverters/buffers and their propagation delays. 
To internally measure a RO frequency, its output is used to trigger a binary counter circuit. 
This counter value is sampled at a user-defined rate, such that changes in the counter value between successive samples of the same RO reflect changes in operating conditions.

\textbf{Delay-Line Based Sensor.}
Another prominent on-chip sensor is a delay line, i.e., a chain of buffers, through which a clock signal propagates. 
Propagation delay is measured by how far the clock propagates through the enabled/open buffers within a fixed time frame using a time-to-digital convertor (TDC) circuit. 
This numeric value at the TDC output is sampled at some user-defined rate, such that changes in TDC value between successive samples reflect changes in operating conditions.

Zhao and Suh~\cite{ZhaSuh18} examined the trade-off between RO and delay-line based sensors in terms of sampling frequency, accuracy, resolution and complexity. 

\subsection{Reconfigurable Power Viruses}
\label{sec:pwr_virus}
Examples of on-chip power dissipating circuits are: \textbf{RO-based}, \textbf{FF-based} and \textbf{PIP-based} power viruses. By toggling a large group of such power viruses for enough time, voltage drops (both transient and steady-state drops) can be induced in the PDN. 
\textbf{FF-based} power virus consists of a D flip-flop (D-FF), whose output is connected to the input of an inverter, while the inverter output is fed back to the input of the D-FF~\cite{GOKT16}. Whereas, \textbf{PIP-based} power virus is constructed by leveraging programmable interconnect points (PIPs) in Xilinx FPGAs. These are programmable transistors used to connect inputs and outputs of I/O blocks and CLBs into routing network in the FPGA fabric.
Unused PIPs and wires in the FPGA fabric resemble a high capacitance such that frequently charging and discharging it result in transient voltage drop and overshoot in the PDN~\cite{ZSZF13}.

On-chip power dissipating circuits can be precisely controlled to induce power glitches within a specific time-frame.
Moreover, such power viruses can be activated for long enough time to heat the FPGA.
Such circuits can be leveraged to mount power fault attacks, as well as on-chip noise generators for mitigating side-channel attacks, as we describe later in more detail.

\subsection{Fault-Injection Attacks}
\label{sec:fa}
\textbf{DoS Attacks.} Gnad et al.~\cite{GOT17} deploy thousands of RO-based power viruses to construct a power-hungry circuit. The excessive switching activities of the ROs, in the picosecond regime, create voltage emergencies in the PDN that cannot be compensated by the decoupling capacitors or the voltage regulator. This causes a Xilinx Virtex 6 FPGA to crash in less than 1 ms. Such a DoS attack is deemed feasible on FPGAs deployed in Amazon EC2 F1 instances~\cite{FPGA2020_tutorial}. 
Another way to induce such DoS attack is by invoking short circuits~\cite{shortcircuits}.  

\textbf{Fault Attacks.} Mahmoud et al.~\cite{MahSto19} also leverage a power-hungry circuit to induce timing failures during the operation of a true random number generator, used for the generation of secret keys, on a Xilinx Virtex 7. The faulty output was shown to be biased and failed randomness tests.
Krautter et al.~\cite{KGT18} show that by precisely controlling a power-hungry circuit it is feasible to induce timing failures in the $9^{th}$ round of a \emph{soft} AES-128 core, i.e., implemented in the FPGA fabric. The induced timing faults are used to recover the AES key on an Intel Cyclone V SoC by means of a differential fault analysis~\cite{PirQui03}, where the victim AES core operated at 111 MHz and spatially shared the FPGA with an array of RO-based power viruses. The achieved key recovery rate is of 93\% on average using an average of 18K encrypted messages.   
The aforementioned attacks require implanting \emph{malicious} logic on the \emph{same} FPGA where the cryptographic cores are running. Weissman et al.~\cite{jackhammer} demonstrate stealthier fault injection attacks, by leveraging the Rowhammer effect~\footnote{Inducing bit flips in DRAM cells by repeatedly accessing their neighboring rows~\cite{rowhammer}.}, that can be mounted from the FPGA on the host's memory subsystem without the need for malicious logic.

\subsection{Side-Channel Attacks}
\label{sec:sca}
Remote side-channel attacks on cloud FPGAs can be classified in two major categories: attacks based on either the \emph{power leakage} or the \emph{crosstalk effect} of the victim circuit. The crosstalk effect is caused by the electromagnetic interference that occurs between unshielded long wires that are laid in parallel (see~\autoref{fig:fpga}). Thus, a signal transmitted through one long wire influences the propagation delay of signals in neighboring long wires (with a maximum distance of two wires). 

\textbf{Power Leakage.} Schellenberg et al.~\cite{SGMT18} deploy a delay-line sensor to detect voltage fluctuations in the PDN caused by the activity of a soft AES-128 core. Using correlation power analysis (CPA), encrypted messages and the measured sensor values, i.e., power traces, the secret key can be fully recovered. The attack is shown to be successful for different locations of the sensor with reference to that of the AES core, with fewer traces required when the sensor is closer to the AES core.

In~\cite{ZhaSuh18}, RO-based voltage sensors are leveraged to monitor the activity of a co-located RSA-1024 core on the same FPGA fabric. Using simple power analysis (SPA) and sampled power traces of 20 RO-based voltage sensors, RSA private key can recovered using an average of 9 encrypted messages, when both victim and attacker circuits are isolated by passive fences of unused logic~\cite{HBWS07}. 
The attack also works when the RSA encryption runs as a process on a CPU that shares the same PDN with the FPGA, i.e., in FPGA-based SoC. The RSA key can also be recovered with more traces/encrypted messages, when another power-consuming circuit is running, whose power consumption is comparable to that of the RSA core.

Such attacks have been also shown possible even when the sensor and the victim logic run on two different FPGA chips integrated on the same board that share the power supply in~\cite{SGMT18e}. \autoref{tab:sca_tab1} summarizes the details of the attacks and the number of required encryption/decryption operations.

\begin{table*}[htbp]
\caption{Remote side-channel attacks leveraging power leakage.}
\begin{center}
\begin{tabular}[c]{|c|c|c|c|c|c|c|c|}
\hline
\multirow{3}{*}{Ref.} & \multicolumn{2}{|c|}{Victim Circuit} & \multicolumn{2}{|c|}{Sensor Circuit} & Required & \multirow{3}{*}{Analysis} & \multirow{3}{*}{\makecell{Proximity between Sensor \& Victim}}\\
\cline{2-5}
 & \multirow{2}{*}{Key Size} & \multirow{2}{*}{Frequency} & \multirow{2}{*}{Type} & \multirow{2}{*}{\makecell{Sampling \\ Frequency}} & \multirow{2}{*}{\makecell{Encrypted \\ Messages}} & & \\
& & & & & & & \\
\hline 
\cite{SGMT18} & \multirow{2}{*}{\makecell{AES (128-bit)}} & \multirow{2}{*}{\makecell{24 MHz}} & \multirow{2}{*}{\makecell{Delay-line}} & 24 $^{\mathrm{a}}$ MHz & 5K & CPA & Single Xilinx Spartan 6 chip ((SAKURA-G board))\\
\cline{5-8}
\cite{SGMT18e} & & & & 24 MHz & 2.5 million & CPA & Two Xilinx Spartan 6 chips (SAKURA-G board)\\
\hline
\cite{ZhaSuh18} & RSA (1024-bit) & 20 MHz & RO & 0.1 MHz & 8.9 $^{\mathrm{b}}$ & SPA & Single Xilinx Artix 7 (Zynq-7020 SoC, Zedboard)\\
\hline
\cite{SGMT18e} & RSA (224-bit) & 24 MHz & Delay-line & 24 MHz & not reported & SPA & Two Xilinx Spartan 6 chips (SAKURA-G board)\\
\hline
\multicolumn{8}{l}{ }\\
\multicolumn{8}{l}{$^{\mathrm{a}}$ Results apply also to other sampling frequencies: 48, 72 \& 96 MHz.}\\
\multicolumn{8}{l}{$^{\mathrm{b}}$ Traces are reported for the case when the RSA and sensor logic are isolated by a passive fence of unused logic~\cite{HBWS07}.}

\end{tabular}
\label{tab:sca_tab1}
\end{center}
\end{table*}

\textbf{Crosstalk Effect.} Ramesh et al.~\cite{RPDP18} exploit the crosstalk effect between long wires in FPGAs to successfully obtain the secret key of a soft AES-128 core.
The transmitter, routed through long wires, is an input to an AES S-box. 
The snooping circuit comprises a RO-based voltage sensor (as in~\autoref{fig:ro2}) triggering a counter. The receiver, which is one of the RO's wires, is routed through long wires and is located next to the transmitter wire.
Using differential power analysis (DPA)~\cite{dpa11}, the counter values obtained from the snooping circuit (power traces) and the encrypted messages, the key of the final round of the AES-128 can be recovered byte by byte. Then the encryption key can be obtained by inverting the key schedule. The crosstalk effect increases for longer transmission pair and lower operating frequencies of the victim logic, which directly influences the number of encryption operations required to recover the key. The attack is evaluated on several boards: Altera DE2-115, Cyclone IV, Stratix V and DE5a-Net Arria 10 GX boards~\cite{RPDP18,PRPE19}. 

The same effect is also demonstrated (albeit in another context) in~\cite{GRE18}, where the authors leverage a sliding-window sampling scheme to sample the RO-based sensor at overlapping time periods in order to extract the value of each transmitted bit of a secret key. 
The authors compute the probability of correctly recovering the key. Their formula indicates that the probability is higher for larger keys (asymmetric vs. symmetric keys) and smaller window sizes, which is demonstrated for 32-bit keys in~\cite{GER19}. 
In \autoref{tab:sca_tab2} we summarize the performance of each of the aforementioned DPA and sliding window approaches. 

Long wire leakage is characterized for different generations and families of Xilinx FPGA devices including Virtex 5, Virtex 6, Artix 7, Spartan 7 and Virtex UltraScale+~\cite{GER19,GRS19e}. 
In particular, long wire leakage in Xilinx UltraScale+ FPGAs, which are deployed in real-world cloud infrastructures, e.g., Amazon F1 and Huawei FP1 instances, can still be measured, even though on Amazon F1 instances ROs with combinatorial loops are prevented in user logic. This is made possible by deploying latch-based RO or flip-flop based RO sensors, shown in~\autoref{fig:ro3} and ~\autoref{fig:ro4}, which bypass detection checks on commercial clouds~\cite{SSNS19}.
More technical details on the crosstalk effect and the attacks can be found in Appendix~\autoref{app:crosstalk}.

\raggedbottom
\begin{table*}[htbp]
\caption{Remote side-channel attacks leveraging crosstalk effect.}
\begin{center}
\begin{tabular}[c]{|c|c|c|c|c|c|c|c|}
\hline
\multirow{3}{*}{Ref.} & \multicolumn{3}{|c|}{Victim Circuit} & \multicolumn{2}{|c|}{Snooping Circuit} & \multirow{3}{*}{Analysis} & \multirow{3}{*}{Platform}\\
\cline{2-6}
& \multirow{2}{*}{Key Size} & \multirow{2}{*}{Frequency} & Transmitter & Sampling & Required & & \\
& & & Length & Frequency & Encrypted Messages & & \\
\hline 
\cite{RPDP18} & AES $^{\mathrm{a}}$ (128-bit) & 1 MHz & \makecell{1-segment\\long wire} & 1 MHz & 233K & DPA & \makecell{Intel Cyclone IV GX \\ (Development Kit)}\\
\hline
\cite{GER19} & 32-bit & 0.78 MHz & \makecell{$\geq$ 1-segment\\long wire} & 195 KHz & 20K $^{\mathrm{b}}$ & \makecell{Sliding Window\\($w = 4$)} & \makecell{Xilinx Artix 7 \\ (Nexys 4 DDR)}\\
\hline
\multicolumn{7}{l}{ }\\
\multicolumn{7}{l}{$^{\mathrm{a}}$ Reported results are for automatically placed and routed victim AES.}\\
\multicolumn{7}{l}{$^{\mathrm{b}}$ For all four counters, i.e., 5K per counter, to recover only 98.4\% of the 32-bit key.}
\end{tabular}
\label{tab:sca_tab2}
\end{center}
\end{table*}

To summarize, exploiting the crosstalk effect requires more power traces and/or encryption operations than a power side-channel attack, as seen in~\autoref{tab:sca_tab1} and~\autoref{tab:sca_tab2}. Furthermore, the crosstalk effect diminishes for victim circuits operating at higher frequencies~\cite{RPDP18}. Besides, it requires close proximity between the victim and snooping circuit with a maximum of one long wire separating the transmitter and the receiver. Thus, it is, in practice, harder to achieve in real-world settings. 

\subsection{Covert-Channel Attacks}
\label{sec:cca}
Known covert channels on cloud FPGAs can be established based on one of three effects, namely crosstalk effect, power leakage and thermal leakage. The first two effects require that both the transmitter and receiver are co-located on the same cloud FPGA at the same time (spatial multi-tenancy) which is not yet adopted by industry. However, thermal leakage can be exploited in real-world cloud FPGAs deployment model, where each client rents a whole FPGA at his/her disposal (temporal multi-tenancy). However, establishing the thermal covert channels between cloud FPGA users requires a pre-agreement on the targeted specific FPGA, which implies disclosing information about the cloud infrastructure, i.e., identifying FPGA instances allocated to tenants.
One approach to acquire such information is by fingerprinting cloud FPGAs using physically unclonable functions to measure unique characteristics of the FPGAs. Tian et al.~\cite{TXGR20} conduct the experiments on Amazon AWS cloud and demonstrate the feasibility of renting the same FPGA by two successive clients for the different AWS instances.

\textbf{Crosstalk Effect.} Besides side-channel attacks, the crosstalk effect can be exploited to construct a covert channel between two circuits sharing the FPGA resources at the same time. The authors in~\cite{GRE18,GER19} propose to use a transmission pair of 2-segment long wire and a transmission period of 82 $\mu$s per bit. Thus, using the Manchester encoding scheme, in which one bit of information is represented as a tuple: `1' as (1,0) and `0' as (0,1), the constructed channel has a bandwidth of 6.1kbps and accuracy of 90 - 99.9\% without using error-correcting codes.
The covert channels have been shown on several Xilinx devices.
 
\textbf{Power Leakage.} By leveraging a grid of RO-based power viruses as transmitters and RO-based voltage sensors as receivers, the authors demonstrate in~\cite{GRS19} a covert channel between two users spatially sharing a cloud FPGA.
Although multi-tenant spatially-shared FPGAs are not yet deployed in real-world cloud solutions, such covert channels are shown on Amazon F1 and Huawei FP1 instances, which deploy Xilinx \emph{3D} Viretx UltraScale+ FPGAs. Virtex UltraScale+ FPGA chips accommodate up to four dies called super logic regions (SLRs) in a single chip to deliver high logic density~\cite{ds890}, where these dies share power and clock delivery systems.
The authors envisioned, in a multi-tenant scenario, that each SLR would be dedicated to a different tenant and demonstrated the construction of a covert channel between two SLRs. 
The covert channel constructed had a bandwidth of 781Kbps (488Kbps for FPGAs operating at a lower frequency)  and an accuracy of 97 - 99.9\% on Huawei FP1 (Amazon F1) boards, assuming all transmitters are enabled and disabled simultaneously and a standard Manchester encoding is deployed. 
Power leakage based covert-channel is further improved and demonstrated on three realistic setups of FPGA-to-FPGA, CPU-to-FPGA and GPU-to-FPGA, where transmitters and receivers are separated on two distinct boards sharing only the external power supply unit in~\cite{GRS20}.

\textbf{Thermal Leakage.} By leveraging a grid of ROs as transmitters (heaters) and RO-based thermal sensors as receivers, the authors in~\cite{TiaSze19} demonstrate building a covert channel between two users temporally sharing a cloud FPGA. A transmitter circuit is configured on the FPGA and enabled to achieve a steady-state temperature corresponding to logic `1'. Then the transmitter circuit is disabled and the FPGA is left idle for a while before configuring the receiver's bitstream on the same FPGA. 
The receiver circuit then measures the ROs frequencies and compares them to nominal ROs frequencies which correspond to logical `0' to acquire the transmitted value. 
The covert channel bandwidth is defined by the heating period, the maximum time gap between heating the FPGA and sensing its temperature, and the transmitter's size (number of ROs). The evaluation is performed on cloud FPGA instances (Intel Stratix V FPGAs), where for an idle period between two users of a cloud FPGA of 120 seconds, 6.64 minutes are required to leak one bit.
Unlike covert channels based on crosstalk or power leakage, thermal leakage impose no constraints on the placements of transmitter and receiver circuits since the generated heat can be sensed from anywhere on the FPGA.
\autoref{tab:cca_tab} shows the different covert channels that can be established in FPGAs and their bandwidths.

\begin{table*}[ht]
\caption{Remote covert-channel attacks.}
\begin{center}
\begin{tabular}[c]{|c|c|c|c|c|c|c|}
\hline
Ref. & Deployed Effect & Coding Scheme & Bandwidth & Error Rate & Multi-Tenancy & Platform\\
\hline
\cite{GER19} & Crosstalk & Manchester$^{\mathrm{a}}$ & 6.1 Kbps & 0.1 - 1\% & Spatial & \makecell{Transmitter \& Receiver both on\\ Xilinx Virtex 5 (ML509),\\Virtex 6(ML605), Artix 7 (ArtyS7),\\Spartan 7 (Nexys 4 DDR \& Basys 3)}\\
\hline
\cite{GRS19} & Power Leakage & Manchester & 488 Kbps & 0.1\% & Spatial & \makecell{Transmitter \& Receiver on\\different SLRs of Xilinx Virtex\\UltraScale+ (Amazon F1)}\\
\hline
\cite{GRS19} & Power Leakage & Manchester & 781 Kbps & 0.1\% & Spatial & \makecell{Transmitter \& Receiver on\\different SLRs of Xilinx Virtex\\UltraScale+  (Huawei FP1)}\\
\hline
\cite{GRS20} & Power Leakage & Manchester & 6.1 bps & $\leq$ 10\% & Temporal & \makecell{Transmitter: Xilinx Kintex 7 (KC705)\\ Receiver: Xilinx Artix 7 (AC701)} \\
\hline
\cite{GRS20} & Power Leakage & Manchester & 6.1 bps & 3\% & - & \makecell{Transmitter: CPU (14 threads)\\ Receiver: Xilinx Artix 7 (AC701)} \\
\hline
\cite{GRS20} & Power Leakage & Manchester & 2 bps & 3\% & - & \makecell{Transmitter: GPU\\ Receiver: Xilinx Artix 7 (AC701)} \\
\hline
\cite{TiaSze19} & Thermal Leakage & none & 1 bit per 6.64 minute & 0\% & Temporal & Intel Stratix V (TACC)\\
\hline
\multicolumn{6}{l}{}\\
\multicolumn{6}{l}{$^{\mathrm{a}}$ Assuming no changes in operating conditions, e.g., monotonic changes in temperature.}\\

\end{tabular}
\label{tab:cca_tab}
\end{center}
\end{table*}

\section{Mitigating Attacks}
\label{sec:def}
We outline next defenses that have been recently proposed to mitigate the attacks described in~\autoref{sec:pla}, which can be classified into either \emph{detection} or \emph{prevention}.

\subsection{Detection of Attacks}
We refer here to runtime detection solutions that aim to identify malicious logic (during its execution) and can thus mainly defend against active attackers.

\textbf{Detection of Fault-Injection Attacks.}
Provelengios et al.~\cite{PHT19} propose to deploy a network of RO-based sensors in the shell of a multi-tenant FPGA to detect voltage drops and locate the center of a power wasting logic. 
This information can be used by CSP to revoke the malicious logic from the FPGA. However, it is likely that, until the malicious logic is located and revoked, timing errors in neighboring logic would still occur, besides generating false positives.

\subsection{Prevention of Attacks}

\textbf{Prevention of Malicious Primitives.}
Such techniques attempt to verify that a user design logic is harmless. Two approaches have been proposed: either applying design rule checks to verify that a netlist does not violate a set of pre-defined rules, or checking the final bitstream by ``antivirus'' programs.  
Amazon AWS enforces design rule checks on the clients' uploaded designs to prevent combinatorial loops, e.g. ROs, before generating the final bitstreams~\footnote{\url{https://github.com/aws/aws-fpga/blob/master/ERRATA.md}}. However, alternative designs (e.g., \emph{sequential} RO circuits) can still bypass these checks~\cite{ZSZF13,GRS19e,SSNS19}.
In~\cite{KGT19,LMGP20} virus scanners are proposed for deployment by the CSP to detect malicious primitives (~\autoref{sec:sensors} and~\autoref{sec:pwr_virus}).
However, reverse-engineering of the bitstream is required to search for uncommon structural and behavioral properties of malicious logic in the recovered netlist.
Nevertheless, this requires regularly updating the virus scanner framework with formulation of properties, i.e., virus signatures, to account for new malicious logic designs.
Most importantly, this approach violates IP protection, although not adopted by industrial cloud services yet, since access to plain text bitstreams is inevitable.

\textbf{Prevention of Crosstalk-Effect Attacks.} Giechaskiel et al.~\cite{GRE18,GER19} propose to add guarding wires, which are the neighboring four long wires, two from each side, of a long victim wire. The guarding wires can be left unoccupied or further driven by random values.
This can be performed either by manual inspection of the placed and routed design to identify the source of leakage and add the guarding wires or by the support of FPGA vendor toolchains. This requires including directives in the HDL design by the developer to annotate sensitive signals and further modifying routing algorithms to automate adding the guarding wires.

\textbf{Mitigation of Power-Leakage Attacks.} Krautter et al.\cite{KGSM19} propose to reduce side-channel leakage by implanting a fence of ROs, which works as a noise generator circuit, around the logic to be protected, thus reducing signal-to-noise-ratio (SNR). The authors also show that a fence activation strategy based on internal power measurements, using feedback measurements from voltage sensors, is more effective than a random activation pattern. This is in line with the findings in~\cite{ZhaSuh18}, where the authors demonstrate the limitation of their attack with co-existing noise. 
However, this approach is both technology- and application-dependent, and requires experimental evaluation on the targeted FPGA board to determine the total amount of fence ROs required to neutralize the voltage fluctuations induced by the cryptographic primitives. 
Other known countermeasures against power leakage or power analysis attacks are hiding and masking. Masking techniques for reconfigurable platforms are either inefficient with respect to area and performance~\cite{MorMis13}, or still generate exploitable leakage~\cite{CHS09}. Moreover, masking techniques should be designed for every cryptographic function individually and the desired security guarantees. 
Hiding techniques on FPGAs suffer well-known issues that cause information leakage. These are imbalanced routing~\cite{NBDD10}, early propagation effect~\cite{SuzSae06} and glitches~\cite{MPO05}.

\textbf{Prevention of Thermal-Leakage Attacks.} A CSP can enforce a minimum idle period, on the order of minutes, after a client releases the FPGA allowing it to recover to a nominal temperature before allocating it to another client~\cite{TiaSze19}. Alternatively, the FPGA can be heated or cooled to a pre-defined temperature before the next client can use the FPGA.

\section{Discussions and Recommendations}
\label{sec:disc}
\subsection{The Future of FPGA Cloud Computing}
FPGA devices have only become increasingly deployed by cloud service providers very recently, but spatial multi-tenant FPGA computing %
is currently not yet supported. However, it is more recently being investigated and considered for deployment. In light of our survey of the state of the art on that front, as well as our analysis of the additional security and privacy concerns that would arise for this cloud FPGA computing deployment, we present next some of our insights and recommendations for the future of FPGA-based cloud computing.

\textbf{Cross-Stakeholder Efforts.} Different stakeholders (outlined in~\autoref{sec:security}) are involved in FPGA cloud computing, each having different concerns, requirements, rights and responsibilities from a security/privacy standpoint. Many of these influence each other and also influence the entire supply chain and deployment process, even prompting low-end requirements on the FPGA fabrication itself (with cloud computing deployment in mind). 
It is thus imperative for relevant stakeholders to continue to actively and systematically analyze and address these concerns and requirements together early on in a coherent and proactive approach.
Through an open and systematic collaboration, FPGA vendors become aware of the requirements of FPGA devices specifically for cloud deployment, since this influences the logic and architectural constructs that would be hard-wired in the device fabric at production. 
Various security solutions proposed by academics in recent years targeting FPGA cloud computing would be significantly more efficient or even only feasible if certain primitives are already incorporated into the FPGA fabric at production. 
Furthermore, for key provisioning and IP protection, FPGA vendor support and involvement is further required after the FPGA device is acquired by the CSP.
Thus, the requirements on FPGA devices used by single users are not the same for devices deployed in the cloud, which some FPGA vendors are already recognizing and addressing with their more recent FPGA devices. 

Similarly, toolchains are required to cater for cloud deployment requirements to provide additional security guarantees, e.g., by extending them with particular design rule checks to enforce certain constraints on routing (e.g., to eliminate potential crosstalk across co-located tenants) and checking for potentially malicious design primitives. The deployment of these toolchains and SDKs, i.e., whether they are at the client's end or hosted by the CSP, and whether they can be trusted, and how justified are these trust assumptions, and consequently what are the security guarantees they can provide, are more open questions that can only be answered through a thorough and open exchange across all stakeholders. 
One way this can achieved more systematically and openly is through a more formal collaboration, such as the Confidential Computing Consortium~\cite{ccc} recently established to standardize the adoption of Trusted Execution Environment (TEE) technologies among industry players. This already includes various CSPs and an FPGA vendor as members, and may be a good starting point to initiate dedicated focus and analysis of FPGAs in cloud computing and establishing relevant standards.

\textbf{Key Provisioning and Management.} Among the various outstanding issues with cloud FPGA computing, IP protection for different FPGA tenants, as well as the required key provisioning and management to guarantee that, are the most prominent and evolving concerns in FPGA-based computing. Being a topic of concern in earlier years for end-users, FPGA vendors have reacted already, as evident in more recent FPGA boards that provide both bitstream encryption and integrity protection as we have presented in~\autoref{sec:ipencrypt}. A user can program his/her secret key into the FPGA in a secure environment and then deploy it in field ready for future secure updates. However, in the cloud scenario, a client has no physical access to the FPGA and thus no secure means to program a secret key remotely into the FPGA. 
Moreover, multiple clients might be allocated to use the same FPGA instance within a relatively short time-frame. Therefore, a remote secret key provisioning and management scheme, tailored for these deployment and usage settings, is required. 
In this direction several proposals have been introduced. They either rely on public key cryptography~\cite{EguVen12,HKKT17} or physically unclonable functions, i.e., PUF-based secret key generation~\cite{EAA19}. Both of which require the support of FPGA vendors, both architecturally and in the key management scheme that follows after the FPGA is acquired by the CSP. 
FPGA vendors, such as Intel and Microsemi, have already began addressing these challenges in their most recent FPGAs, e.g., Intel Stratix 10~\cite{intel_sdm,intel_device_security} and Microsemi SmartFusion 2~\cite{ug0443}, but key management to multiple cloud clients still remains unaddressed.

\textbf{Lessons Learned.}
We can draw various lessons from CPU-based cloud computing that would analogously translate to FPGA-based cloud computing.
One most critical open question is the means to establish a trust anchor on an FPGA in the cloud, analogous to CPU-based TEEs~\cite{sgx,amd-sev}. A supporting infrastructure (architecturally in FPGA devices as well as within overlying protocols and toolchains provided, besides support from relevant stakeholders) is required to provide the client with means to exchange a secret session key with the FPGA trust anchor, securely communicate and provision his/her IP on the FPGA, and to have sufficient guarantees that this IP (and all processed data) remains contained in secure and untampered isolation. 
Moreover, establishing trust in cloud FPGAs would consequently enable the integrity and confidentiality of data processed by the FPGA and stored in external DRAM memory (assuming physical access adversaries), which is yet another security challenge for trusted computing. On cloud FPGAs, this can only be enabled once secure exchange of clients' secret keys for data encryption and authentication is supported, thus mitigating potential sniffing, spoofing, splicing and replay attacks on external DRAM by CSPs.
 
This, further, calls for a comprehensive analysis of the minimal trusted computing base (both in hardware and software) required to establish the trust anchor and provide the required security/privacy guarantees. This includes investigating efficient secure key and certificate provisioning, storage, and readout, as well as the minimal hardware primitives required to establish an FPGA-based root of trust (cryptographic primitives, their formally verified implementations, side-channel resilience, etc.). Furthermore, which of these primitives need to be ``baked in'' or hard-wired into the FPGA fabric and which can be provided as configurable logic? How much hardware-software co-design is needed and what are the ideal design choices therein? How can debug interfaces, such as JTAG among others, be provided securely to still enable cloud clients to debug their FPGA development without threatening the confidentiality of the IP and computations running on the FPGAs? Various usability/performance/security trade-offs are involved within these open questions, all of which require a comprehensive and systematic treatment, and practical solutions by FPGA vendors, that would be adopted by commercial CSPs. 

Only in recent years, did researchers and practitioners uncover security flaws in microachitectural optimizations, e.g., caches and speculative execution, that have driven our modern processors for decades. More of these flaws originate in the hardware and can only be fundamentally patched by changing the hardware. More critically, some of these flaws undermine the security guarantees of TEEs established in these processors. Many valuable lessons can be drawn from this for cloud-based FPGA computing; a rigorous and systematic security analysis of the underlying fabric and trust anchor is crucial for the derived security guarantees of all that lies on top of that. Furthermore, for FPGAs, unlike CPUs, the raw hardware fabric itself is exposed to users to reconfigure and customize with complete freedom, which poses intuitively an even larger attack surface. In multi-tenant FPGA computing, when the hardware fabric is fundamentally shared by multiple tenants operating on the same device, the same power delivery networks are also shared. Thus, complete isolation at the physical level is fundamentally not possible, and there will always be means to induce malicious physical effects, so long as tenants are free to arbitrarily program their user logic. As shown in~\autoref{sec:def}, there is no foolproof way to vet and police the provided bitstreams for such malicious circuits or their signatures.

Also analogously to TEE infrastructures and SDKs in CPU-based computing, trust in FPGA SDKs is to be carefully examined and trust assumptions are to be questioned and justified, even when these SDKs are provided by FPGA vendors and CSPs. The critical security implications of a memory corruption vulnerability, as found in the SGX SDK, have been demonstrated in~\cite{guard-dilemma}, where the authors show how this vulnerability is exploited to craft a return-oriented-programming attack chain in the Trusted Runtime System (tRTS) code handling enclave entry.
Similarly, FPGA vendor and CSP SDKs for cloud development need to be scrutinized and tested for such vulnerabilities and hardened accordingly.

\subsection{Call to Action}
Given the aforementioned open challenges and research questions in deploying end-to-end secure multi-tenant FPGA cloud computing, we conclude with a call for future research and investigation by both academia and all relevant stakeholders collaboratively in industry.
While academics have shown and reported a spectrum of remotely-exploitable FPGA-based physical attacks, which demonstrate the threats of FPGA multi-tenancy, these actually only touch one dimension of the problem. This still leaves other, perhaps even more fundamental, dimensions uncharted and unaddressed in the end-to-end deployment process, which call for immediate attention and action if FPGA-based cloud computing would be much more ubiquitously adopted by enterprises.

\textbf{Open Standards, Documentation and Disclaimers.}
Resulting from an open collaborative exchange between all stakeholders involved (ideally through a formal consortium), open standards and certifications of how TEEs can be adopted for FPGA-based computing can be established and enforced for all relevant entities. This would provide mutual guidelines for CSPs and FPGA vendors on their provided devices and services, and would equally reassure clients with certain and clear security and privacy guarantees.  
Along the same lines, CSPs and FPGA vendors would be able to document more openly and publicly their security settings and assumptions. This would provide better client education, awareness and disclaimers on how much security guarantees are, indeed, provided in terms of IP and data protection, and what are the configuration choices available, perhaps, for a client. 

\textbf{Spatial Multi-Tenant FPGA Computing.}
Remote physical attacks, surveyed in~\autoref{sec:pla}, preemptively demonstrate and emphasize why spatial multi-tenant FPGA computing is discouraged, or at best adopted cautiously with additional vetting applied to the provided bitstreams by the co-tenants.  
On the other hand, while solutions for detection of malicious circuit constructions and suspicious behavior are reported in literature, they remain specific fixes to specific attacks. They provide no guarantees for future zero-day attacks and some cannot be applied by reconfiguring logic and require logic baked in by the FPGA vendor. This would then require the CSP to exchange its own FPGA devices with new ones ``patched'' in silicon. Nevertheless, in quest for profitability and as FPGA devices' capacities keep scaling, CSPs may eventually adopt spatial multi-tenancy. If so, tenants must be informed if they are sharing an FPGA instance (along with security disclaimers on what are the likely threats) and should be allowed to select (and thus pay) to acquire a dedicated FPGA instance, if desired.

\section{Conclusion}
\label{sec:conc}
In this paper, we survey industrial and academic deployment models of FPGAs in the cloud computing settings, and highlight their different adversary models and security guarantees, while shedding light on their fundamental shortcomings from a security standpoint. 
We further survey and classify existing academic works on recent remotely-exploitable physical attacks on multi-tenant FPGA devices, where these attacks are launched remotely by malicious clients sharing physical resources with victim clients. Through investigating the problem of end-to-end multi-tenant FPGA deployment more comprehensively, we shed the light on other open security and privacy challenges, e.g., IP protection. We conclude with our insights and a call for future research to tackle these challenges.
\ifCLASSOPTIONcompsoc
  \section*{Acknowledgments}
\else
  \section*{Acknowledgment}
\fi
This work was funded by the Deutsche Forschungsgemeinschaft~(DFG) –- SFB~1119 CROSSING/236615297 and the Intel Collaborative Research Institute for Collaborative Autonomous \& Resilient Systems (ICRI-CARS). 

\bibliographystyle{IEEEtran}
\bibliography{references}

\appendices
\subsection{Ring Oscillator: Different Structures}
\label{app:ro}
\begin{figure}[htbp]
	\centering
  \subfloat[\small Pure combinatorial RO: odd number of inverters/stages.\label{fig:ro1}]{
       \includegraphics[width=\linewidth]{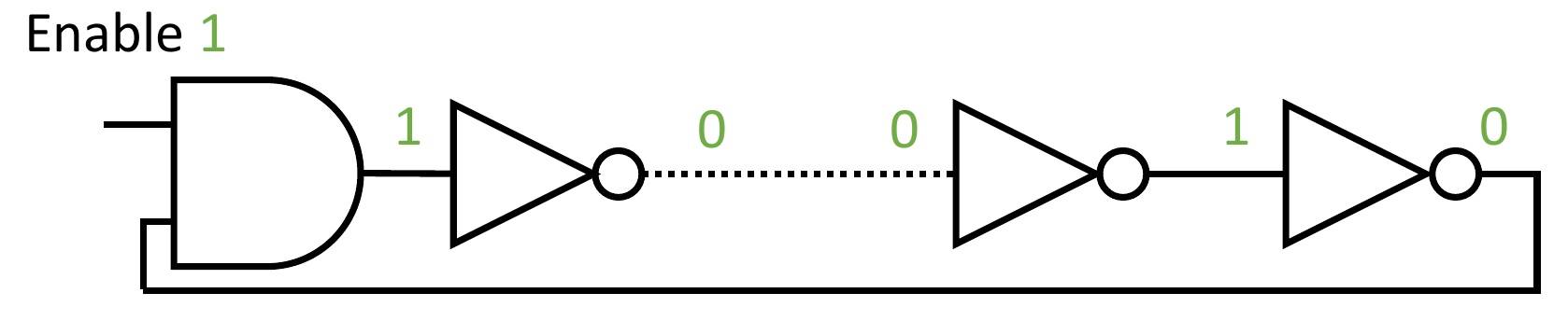}}
    \hfill 
  \subfloat[\small Pure combinatorial RO: one inverter \& even number of buffers.\label{fig:ro2}]{
        \includegraphics[width=\linewidth]{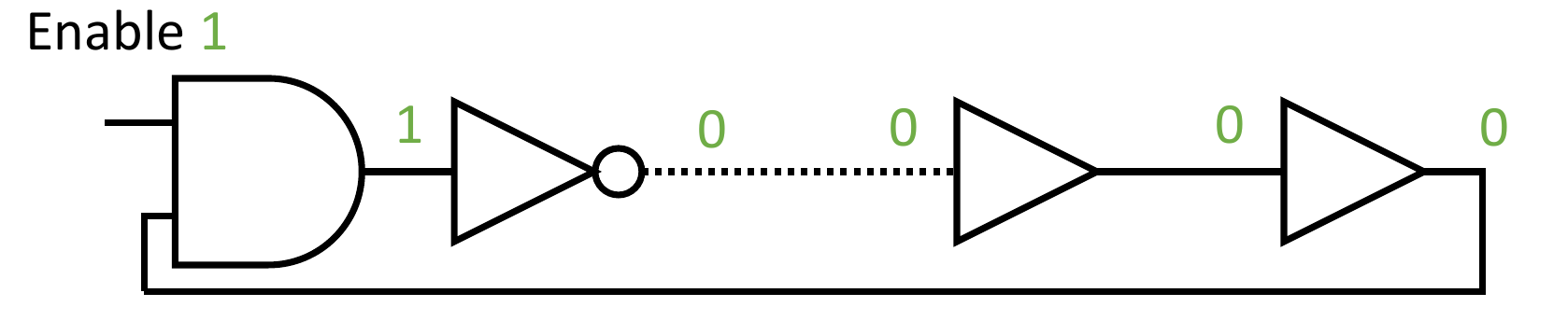}}
		\hfill
  \subfloat[\small Sequential RO design with a D-latch~\cite{GRS19e}.\label{fig:ro3}]{
        \includegraphics[width=\linewidth]{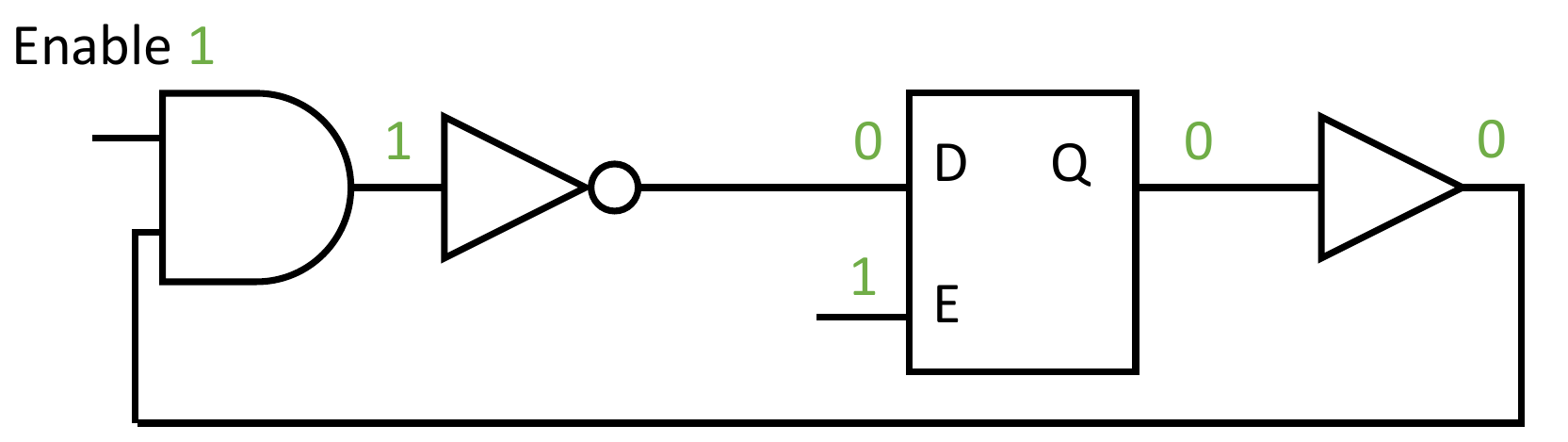}}
    \hfill
  \subfloat[\small Sequential RO design with a D-FF~\cite{GRS19e}.\label{fig:ro4}]{%
        \includegraphics[width=\linewidth]{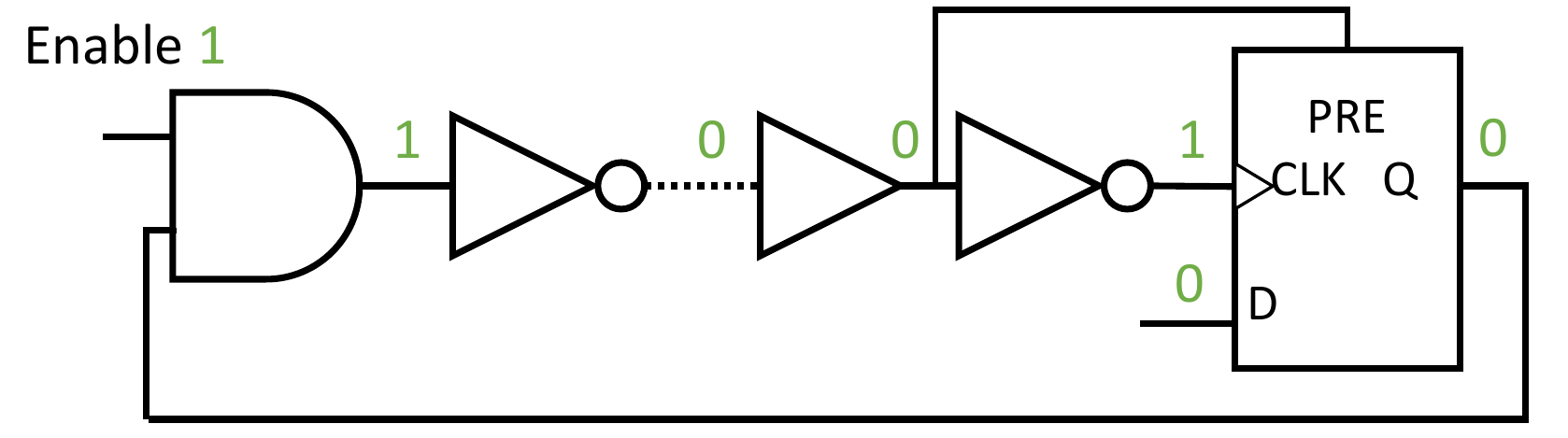}} 
	\caption{\small Different RO Designs.}   
	\label{fig:ro}
\end{figure}

A classical ring oscillator (RO) is shown in \autoref{fig:ro1}. It is also possible to keep only one inverter and replace the rest with digital buffers as shown in~\autoref{fig:ro2}. Both RO designs~\autoref{fig:ro1} and~\autoref{fig:ro2} are \emph{pure combinatorial} logic circuits, whereas the latch-based RO and the flip-flop (FF) based RO, shown in~\autoref{fig:ro3} and~\autoref{fig:ro4} respectively, are sequential. 

\raggedbottom
 
\subsection{Details on Attacks Exploiting the Crosstalk Effect}
\label{app:crosstalk}
\textbf{Attack \cite{RPDP18}.} In the attack by Ramesh et al.~\cite{RPDP18}, the key of the final round of the AES-128 is recovered byte by byte. This requires the attacker and victim circuits on the same FPGA to have 16 transmitter-receiver pairs of neighboring long wires, with one pair for each byte of the final round key, in case of a high-throughput AES implementation with a datapath of 128-bit. Consequently, this is reduced to one pair of neighboring long wires if the AES has an 8-bit datapath. 

The crosstalk effect increases for longer transmission lines and lower operating frequencies of the victim logic, which directly influences the number of encryptions required to recover the key. For example, a minimum of 217 encryptions are required to recover the key for an AES core running at 10KHz, whereas at a higher operating frequency of 4MHz a 1.5 million encryptions are required. Interestingly, this crosstalk effect can also be leveraged to build an adjacency map for long wires within each channel in the FPGA~\cite{PRPE19}, which is considered a proprietary information.

\textbf{Attacks \cite{GRE18} and \cite{GER19}.} The same crosstalk effect is also demonstrated (albeit in another context) in~\cite{GRE18}, i.e., when integrating encrypted or obfuscated third-party FPGA designs from multiple IP vendors in single design. 
In \cite{GRE18} the authors observe that, the propagation delay of the receiver's long wire is highly influenced by how long a neighboring victim long wire carries a signal `1' within a sampling period. Thus, the counter value, i.e., the frequency of the RO connected to the receiver, indicates the hamming weight of the transmitted bits over the victim long wire within that sampling period. Therefore, the authors propose a sliding-window scheme to sample the counter at overlapping periods to extract the value of each transmitted bit of a secret key. By subtracting two overlapped hamming weights, information on the first and last bits of the corresponding windows are revealed.  
Assuming that the victim long wire is transmitting the whole key bits sequentially, and using a sliding-window of size $w$ bits and a secret key of size $N$ bits, where $N$ is a multiple of $w$, i.e., $N=nw$, the authors compute the probability of recovering the key:
\begin{equation}
P={\left(1-\frac{1}{2^{n-1}}\right)}^w  \label{eq}
\end{equation}
This implies that the probability of key recovering is higher for larger keys (asymmetric vs. symmetric keys) and smaller window sizes. 
\autoref{eq}, however, does not take into account noisy hamming weight measurements due to operating conditions (transmission frequency) and layout of transmission pair (length of transmission lines and distance between them).   
The sliding-window approach is applied to 32-bit keys in~\cite{GER19}. The authors demonstrate that using a sliding window of size $w = 4$ bits, 98.4\% of the 32-bit key can be recovered assuming a single bit stays constant on the transmitter for 1.28 $\mu$s, i.e., 780 KHz signal frequency, and a transmission pair of at least 1-segment long wire. Moreover, it requires connecting four counters (equal to the window size) to the snooping circuit, where these counters are sampled at overlapping periods of 5.12 $\mu$s (4 $\times$ 1.28 $\mu$s).

\end{document}